\newcommand*{\HDi}{HD$^{+}$}
\newcommand*{\ddmu}{$dd\mu^{+}$}
\newcommand*{\apHe}[1]{$\bar{p}\,^{#1}$He$^{+}$}
\newcommand*{\mCF}{$\mu$CF}
\newcommand*{\et}{\emph{et al.}}
\def\apj{Astroph.\ J.\ }

\def\jpb{J. Phys.\ B }

\documentclass[aps,pra,twocolumn,groupedaddress,showpacs]{revtex4}
\usepackage{graphicx}
\usepackage{braket}
\usepackage{xfrac}

\begin{document}

\title{
Bounds on fifth forces at the sub-\AA\ length scale}

\author{E. J. Salumbides}
\affiliation{Department of Physics and Astronomy, and LaserLaB, VU University, De Boelelaan 1081, 1081 HV Amsterdam, The Netherlands}
\author{W. Ubachs}
\affiliation{Department of Physics and Astronomy, and LaserLaB, VU University, De Boelelaan 1081, 1081 HV Amsterdam, The Netherlands}
\author{V. I. Korobov}
\affiliation{Joint Institute for Nuclear Research 141980, Dubna, Russia}

\date{\today}

\begin{abstract}
Constraints on a possible fifth-force interaction between hadrons are derived based on an analysis of results from laser precision frequency measurements of antiprotonic helium atoms, both \apHe{4} and \apHe{3} species, and from experiments on resonant formation rates of \ddmu-ions in muon-catalyzed fusion processes.
A comparison is made between accurate experimental data and first-principles theoretical descriptions of the exotic systems within a quantum electrodynamical framework.
The agreement between theory and experiment sets limits on a possible additional hadron-hadron interaction written in the form of a Yukawa potential $V_5(r)=\alpha_5 \exp(-r/\lambda)/r$, with $\lambda$ representing the characteristic length scale associated with the mass of a hypothetical force-carrying particle via $\lambda=\hbar/(m_5c)$.
The laser spectroscopic data of antiprotonic helium set a constraint of $\alpha_5/\alpha_{EM} < 10^{-8}$ for $\lambda < 1$ \AA, while the binding energy of the muonic molecular deuterium ion delivers a constraint of $\alpha_5/\alpha_{EM} \sim 10^{-5}$ for $\lambda < 0.05$ \AA, where $\alpha_{EM}$ represents the strength of the electromagnetic interaction or the fine structure constant.
\end{abstract}

\pacs{12.60.-i, 12.20.-m, 36.10.Gv, 36.10.Ee}

\maketitle

\section{Introduction}

Despite the great success of the Standard Model of physics in describing physical processes at the microscopic scale, it is not considered complete, for it does not encompass gravity.
Furthermore, it lacks a description of Dark Matter and an explanation of the present accelerated expansion of the Universe, which may be associated with Dark Energy or repulsive gravitation. The concept of Dark Matter~\cite{Bertone2005} can alternatively be explained in terms of a deviation from the law of gravity at large length scales, via various modified newtonian dynamics (MOND) theories~\cite{Milgrom1983}. String theory~\cite{Aharony2000} predicts the existence of higher-order dimensions that may be compactified; this compactification is postulated to give rise to deviations from Newtonian gravity at short lengths scales varying from the sub-$\mu$m to the mm scale~\cite{Hamed1998,Antoniadis1998}.
Recent results of laser spectroscopic measurements on muonic hydrogen ($\mu^-$p$^+$)~\cite{Pohl2010,Antognini2013} are in disagreement with similar studies in atomic hydrogen (e$^-$p$^+$) in particular for derived values for the proton size, at the level of 7$\sigma$. These deviations might be ascribed to deviations from quantum electrodynamics (QED), possibly a deviation from Coulomb's law of electromagnetism at short length scales. These examples illustrate the rationale to search for additional forces.

The present study focuses on phenomena in the QED-sector, for which the (Coulomb) interaction potential is given by:
\begin{equation}
    V_{EM}= Z_1Z_2 \frac{\alpha_{EM}}{r} \hbar c ,
\end{equation}
where the coupling strength is the fine structure constant $\alpha=\alpha_{EM}=e^2/4\pi\epsilon_0 \hbar c$.
Deviations from physical law could be expressed as a modification, or in the mathematically equivalent form, as an additional \emph{fifth force}:
\begin{equation}
 V_5=  N_1N_2 \alpha_5 \frac{\exp(-r/\lambda)}{r} \hbar c = N_1N_2 \alpha_5 Y(r) \hbar c,
 \label{eq:Yukawa}
\end{equation}
where the prefactors $N_1$ and $N_2$ could relate to some charge under the fifth force.
In the rest of the discussions we associate $N_{1,2}$ with the hadron numbers, with the expectation that the charge is proportional to the particle number.
The fifth force is parameterized by a generalized Yukawa potential for a certain effective range $\lambda$, which is associated with the mass of a hypothetical bosonic gauge particle of mass $m_5=\hbar/{\lambda c}$, which would act as the force-carrying particle.

The dimensionless coupling constant $\alpha_5$ may be related to the strength of any known interaction, e.g. electromagnetism or gravity.
The latter is represented by:
\begin{equation}
 V_G= N_1N_2 \frac{\alpha_{G}}{r} \hbar c ,
\end{equation}
where $\alpha_G$ is the dimensionless coupling constant for gravity. If we take the proton mass as the mass scaling unit, then $\alpha_G$ can be related to gravitational constant $G$ by the relation $\alpha_G = G m_p^2/(\hbar c)$.
The ratio between the gravitational and electromagnetic coupling constants between two protons is $\alpha_G/\alpha_{EM}=8.1\times10^{-37}$, assuming the inverse-square law behavior of the respective interactions holds.

Tests of the inverse-square law behaviour of gravity have been carried out over an enormous distance scale from kilometers to submicrons, where the latter short-distance constraints are obtained from Casimir-force experiments~\cite{Lamoreaux2012}.
Recently an analysis has been performed on a fifth-force contribution at the typical distance scale of chemical bonds, thus at length scales of 1 \AA. Precision measurements on HD$^+$ ions~\cite{Koelemeij2007,Bressel2012} and H$_2$, D$_2$ and HD neutral molecules~\cite{Liu2009,Salumbides2011,Dickenson2013}
in comparison with advanced QED calculations for the HD$^+$ ion~\cite{Korobov2006,Korobov2008a} and for neutral hydrogen molecules~\cite{Piszcziatowski2009,Komasa2011} allowed for a determination of a constraint $\alpha_5/\alpha_{EM} < 10^{-9}$ at length scales of 1 \AA\ and larger~\cite{Salumbides2013}. We note that analogous constraints for extra \emph{lepton-hadron} interactions may be obtained from a comparison of very accurate experimental and theoretical results on simple atoms and ions \cite{Karshenboim2010}, e.g. H, He, and He$^{+}$.

\begin{figure}
\includegraphics[width=0.9\columnwidth]{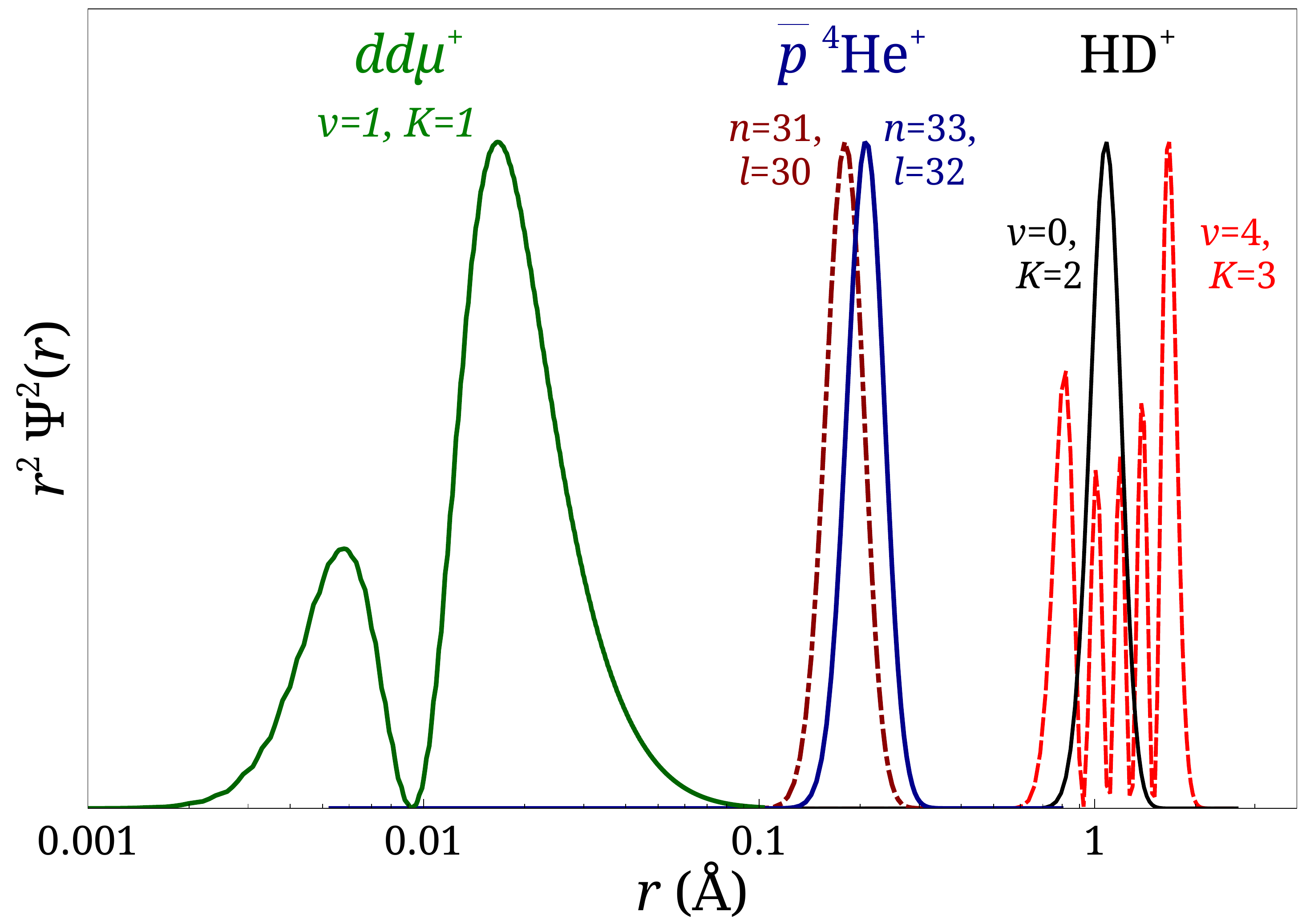}
\caption{
(Color online) The wavefunction densities of relevant states in the systems analyzed in the present study: the weakly bound (1,1) state of $dd\mu^+$ system, two states of antiprotonic-$^{4}$He involved in a measured two-photon transition, and two states involved in $R(2)$ transition the (4,0) band in the \HDi\ ion.
($v,K$: vibrational, rotational quantum numbers)
\label{fig:wfn}}
\end{figure}

In the present study, these results are extended to shorter length scales by considering two exotic atomic or molecular systems. Recent results of laser spectroscopic experiments on antiprotonic helium~\cite{Hori2011} are interpreted in order to derive a constraint on $\alpha_5/\alpha_{EM}$ in the interval $0.05 - 1$ \AA, which is possible due to the smaller separation between heavy particles in these exotic atoms.
In addition, from the binding energy a weakly bound ($v=1, K=1$) state in the \ddmu\ system, determined by temperature-dependent formation rate measurements in muon catalyzed fusion~\cite{Balin2011}, bounds of $\alpha_5/\alpha_{EM}$ in the range $0.005 - 0.01$ \AA\ are derived.
The spatial extent of the wavefunctions, actually $r^2\Psi^2(r)$, for some relevant states in the \ddmu\ and \apHe{}\ systems are plotted in Fig.~\ref{fig:wfn} to indicate the sub-\r{A}ngstrom length-scale accessed. Also drawn in Fig.~\ref{fig:wfn} are the wavefunctions relevant to the tightest constraints obtained from the \HDi\ system.
An assumption is made for the present systems investigated here, similar to that in the analysis which provided constraints from molecules for $\lambda >1$ \AA, that the effects of gravitational, weak and strong interactions do not play a role.
Thus, a comparison between experiment and theory can be made based on calculations solely in the domain of QED.

\section{Antiprotonic helium}

Antiprotonic helium (\apHe{}) is an exotic neutral system composed of a helium nucleus with an antiproton replacing one of the two electrons in a He atom. This long-lived exotic atom, or molecule in view of the heavy interacting particles, was discovered some 20 years ago at the KEK accelerator facility in Japan \cite{Yamazaki1991}.
Antiprotonic substitution takes place when antiprotons are brought to rest in a liquid helium target, where 
almost all antiprotons captured by the helium atom promptly annihilate in the subsequent encounter with the helium nucleus.
A small fraction of the captured antiprotons, in particular, those in states occupying nearly circular orbitals around the He nucleus, is stable against collisions and may survive as long as several microseconds.

The surprising longevity  allows for the manipulation of these \apHe{}\ states, e.g.\ by high precision measurements of laser induced transitions \cite{PhysRep2002}. The accurate measurements of a set of one-photon transitions both in He-3 and He-4 isotopes \cite{Hori2006} were included into the CODATA adjustment of the fundamental physical constants of 2006 and particularly of the (anti)proton-to-electron mass ratio. The fractional measurement accuracy of single-photon laser spectroscopy experiments of \apHe{}, however, is limited by the Doppler effect.

More accurate results have been obtained recently from a Doppler-reduced two-photon laser spectroscopic experiment \cite{Hori2011}.
In order to enhance the two-photon transition probability, two counterpropagating laser beams of slightly unequal frequencies where used, with the frequency of one detuned by some 6 GHz from an intermediate state.
Due to the near-equal frequencies of the counterpropagating beams the first-order Doppler effect largely cancels out, allowing for a more precise spectral line recording, where the hyperfine structure is partially resolved.

\begin{table}
\begin{center}
\caption{A list of contributions to the transitional frequency (in MHz) of the two-photon $(n=36,\ell=34)\to(34,32)$ transition in the antiprotonic helium atom \apHe{4}.
The uncertainty in the first parentheses is the contribution from higher-order terms, while that in the second is due to numerical errors.}
\label{pbarHe:tab1}
\begin{tabular}{l@{\,\,}c@{\hspace{3mm}}r@{.}l}
\hline\hline
$\Delta E_{nr}$        &=&~1\,522\,150\,208&13    \\
$\Delta E_{\alpha^2}$  &=&         -50\,320&63    \\
$\Delta E_{\alpha^3}$  &=&           7\,069&5(0.3)\\
$\Delta E_{\alpha^4}$  &=&              113&1     \\
$\Delta E_{\alpha^5}$  &=&              -11&3(2.1)\\
\hline
$\Delta E_{total}$     &=& 1\,522\,107\,058&8(2.1)(0.3)\\
\hline\hline
\end{tabular}
\end{center}
\end{table}

\begin{table*}
\begin{center}
\caption{Transition frequencies of two-photon $(\Delta n=2, \Delta \ell=2)$ transitions in \apHe{}~\cite{Hori2011}. Theoretical uncertainties are from uncalculated QED terms and numerical errors, respectively. Experimental uncertainties indicate total, statistical, and systematic errors, respectively.
The difference between the experimental and theoretical values as well as the combined experimental-theoretical uncertainty $\delta E$ are also listed.
All values are in MHz.}
\label{pbarHe:tab2}
\begin{tabular}{cc@{\hspace{5mm}}c@{\hspace{5mm}}c@{\hspace{5mm}}r@{\hspace{5mm}}r}
\hline\hline
  & $(n,\ell)\to(n\!-\!2,\ell\!-\!2)$ & Theory & Experiment & Diff & $\delta E$\\
\hline
\apHe{4} & $(36,34)\to(34,32)$ &
     $1\>522\>107\>058.8(2.1)(0.3)$ & $1\>522\>107\>062(4)(3)(2)$ & 3.2& 4\\
                       & $(33,32)\to(31,30)$ &
     $2\>145\>054\>857.9(1.6)(0.3)$ & $2\>145\>054\>858(5)(5)(2)$ & 0.1& 6\\
\apHe{3} & $(35,33)\to(33,31)$ &
     $1\>553\>643\>100.7(2.2)(0.2)$ & $1\>553\>643\>100(7)(7)(3)$ & -0.7& 10\\
\hline\hline
\end{tabular}
\end{center}
\end{table*}

Accurate theoretical results for the three-body $\bar{p}$He$^+$ system where obtained in terms of power series expansion in the fine structure constant $\alpha$.
The nonrelativistic energies were obtained with an accuracy of 16 significant digits by using a variational expansion \cite{KBM99}. Since these states are truly resonant states, the Complex Coordinate Rotation (CCR) approach has been used in order to obtain square integrable wave functions, and the Rayleigh-Schr\"odinger perturbation theory is applied to an isolated CCR state \cite{Simon73}. Details of calculations may be found in \cite{Korobov2008b}. Here we point out that the leading order relativistic corrections in the form of the Breit-Pauli Hamiltonian and the leading-order radiative correction considered account for recoil corrections, while the higher-order terms were taken within the nonrecoil approximation. The finite size corrections were also included, however, their contribution to the \emph{vibrational} or \emph{inter-Rydberg} transitions strongly cancels out, such that the uncertainty contribution of the uncertainty in the nuclear charge radii is negligible. The energy contributions for the $(n=36,\ell=34)\rightarrow(34,32)$ transition in \apHe{4}\ are listed in Table \ref{pbarHe:tab1}. The uncalculated higher-order QED terms as well as numerical errors in the calculations contribute to the total uncertainty in the theoretical transition energies.

The theoretical results and the most accurate experimental data from Hori \et~\cite{Hori2011} on precision two-photon spectroscopy of the antiprotonic helium atoms, both in $^3$He and $^4$He, for the three observed transition are presented in Table \ref{pbarHe:tab2}.
Recent progress in the calculation of the one-loop self-energy contribution of the order $m\alpha^7$ (or $\alpha^5E_{nr}$) should allow for the improvement in the theoretical predictions of transition frequency intervals to a relative uncertainty of $10^{-10}$~\cite{KKH2013}.

The effect of a fifth force is evaluated by treating the potential in Eq.~\ref{eq:Yukawa} as a perturbation on the level energy of the particular states, resulting in a shift in the transition energy of
\begin{eqnarray}
     \left\langle\Delta V_{5,\lambda} \right\rangle  &=&  \alpha_5 N_{1} N_{2} \left[ \left\langle \Psi_{n',\ell'}(r)  |  Y(r,\lambda) | \Psi_{n',\ell'}(r) \right\rangle \right. \nonumber \\
                   && \qquad \left. - \left\langle \Psi_{n'',\ell''}(r)  |  Y(r,\lambda) | \Psi_{n'',\ell''}(r) \right\rangle  \right] \hbar c \nonumber \\
                    &=& \alpha_5 N_{1} N_{2} \Delta Y_{\lambda} \hbar c
\label{YV5}
\end{eqnarray}
between two states represented by the wavefunctions $\Psi_{n',\ell'}(r)$ and $\Psi_{n'',\ell''}(r)$, which express the probability of finding the He$^+$ and $\bar{p}^{-}$ at a certain separation $r$ within the exotic atom. For the \apHe{4} isotope $N_1=4$ and $N_1=3$ for the \apHe{3} isotope, while in both cases $N_2=1$ for the antiproton.
In the experiment, the long-lived antiprotonic helium states have quantum numbers $n\sim\ell\sim30$. The wavefunctions of the ($n=33,\ell=32$) and ($n=31,\ell=30$) states are indicated in Fig.~\ref{fig:wfn} indicating the distance range accessed.
The effect expressed in Eq.~\ref{YV5} for transition energies is differential, and the sensitivity for probing a fifth force $V_5$ is larger if the wavefunctions involved in the transition have different internuclear separations.
Numerical integrations using the wavefunctions from theory were performed to evaluate the $\left\langle\Delta V_{5,\lambda}\right\rangle$ contribution for the transitions of relevance, treating $\lambda$ as a parameter.

The experimental $E_\mathrm{exp}$ and theoretical $E_\mathrm{th}$ transition energies are in good agreement, the difference $E_\mathrm{exp} - E_\mathrm{th}$ being consistent with zero as listed in Table~\ref{pbarHe:tab2}. The combined uncertainty $\delta E=\sqrt{E_\mathrm{exp}^2 + E_\mathrm{th}^2}$ between the experimental and calculated values were used to set bounds for the maximum contribution to a fifth-force. A constraint for the coupling strength $\alpha_5$ is obtained for a range of values of an interaction length $\lambda$ by the relation
\begin{equation}
   \alpha_5 < \frac{\delta E}{N_{1} N_{2}\Delta Y_{\lambda} \hbar c}.
   \label{Eq:constraint}
\end{equation}

\begin{figure}
\includegraphics[width=\columnwidth]{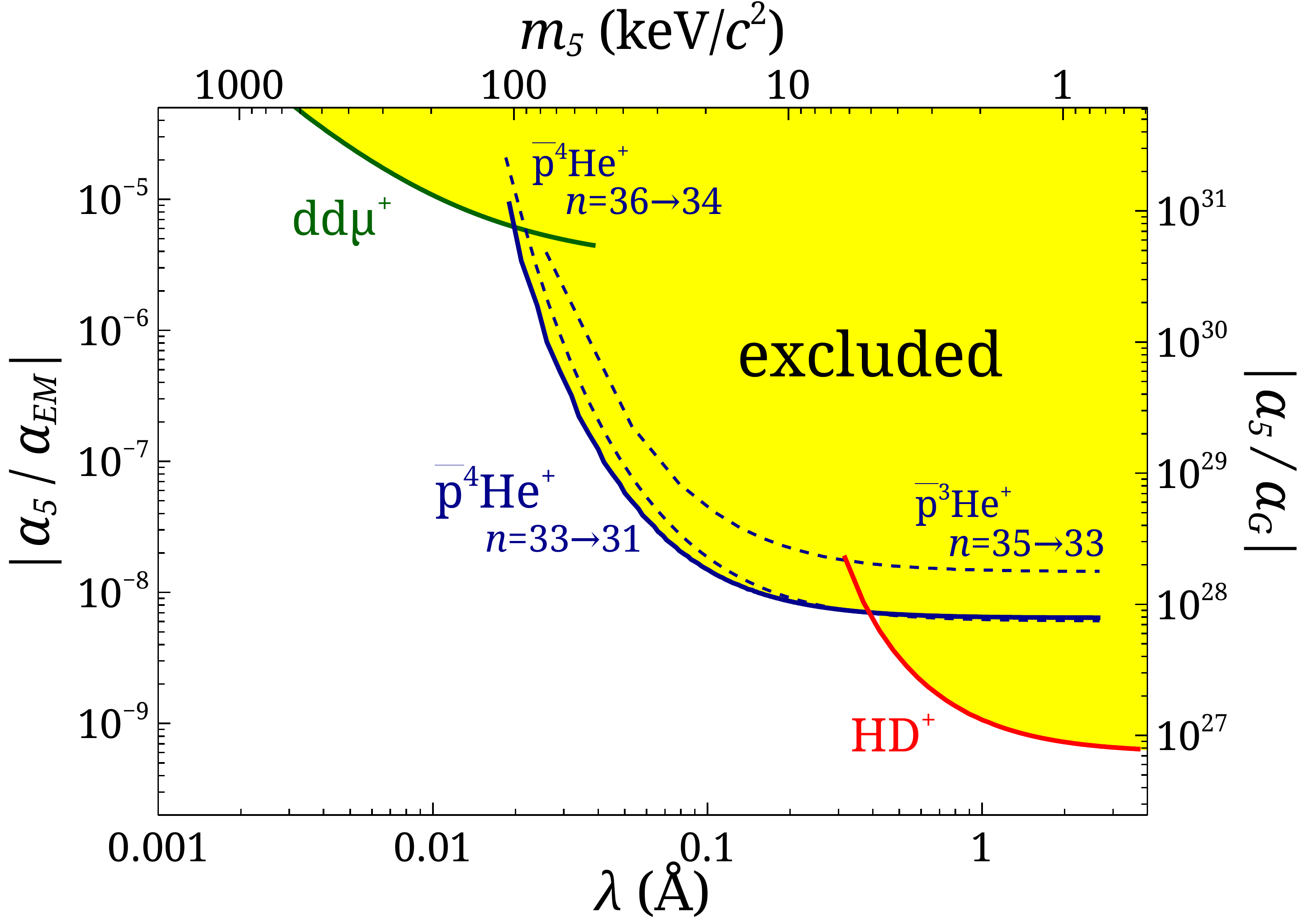}
\caption{
(Color online) Bounds on the coupling strength $\alpha_5$ for a fifth force as derived from the binding energy the weakly-bound $(v=1, K=1)$ state in \ddmu\ from \mCF\ studies, transitions in antiprotonic helium, and from transitions in \HDi~\cite{Salumbides2013}. The coupling strength is compared to $\alpha_{EM}$ (left scale) and $\alpha_G$ (right scale) as a function of fifth-force range $\lambda$. The upper scale gives the mass of the force-carrying particle $m_5$.
\label{expectation}}
\end{figure}

From the experiments reported by Hori et al.~\cite{Hori2011}, two transitions were measured in the \apHe{4}\ isotope: ($n=36\rightarrow34,\ell=34\rightarrow32$) and ($n=33\rightarrow32,\ell=31\rightarrow30$); while the ($n=35\rightarrow33,\ell=33\rightarrow31$) transition was measured in the \apHe{3}\ isotope.
The derived limits for for $\alpha_5$, parametrized for $\lambda$, from both antiprotonic helium isotopes are shown in Fig.~\ref{expectation} for all three transitions.
The limit for $\alpha_5$ obtained from ($n=33\rightarrow32,\ell=31\rightarrow30$) transition in \apHe{4} is drawn as a solid line, where the more penetrating $\Psi(r)$ as indicated in Fig.~\ref{fig:wfn} provides more stringent bounds for shorter $\lambda$ interaction ranges.
For interaction lengths $\lambda < 1$ \AA, a constraint of $\alpha_5 < 7 \times 10^{-9} \alpha_{EM}$ is obtained from antiprotonic helium, however, this is superseded by constraints from \HDi\ ions for $\lambda > 0.4$ \AA.

\section{\ddmu\ molecular ion}

The muonic molecular deuterium molecular ion \ddmu\ is an exotic molecule where a muon $\mu^-$ substitutes the electron in the analogous D$_2^+$ system. Since the muon is about 200 times heavier than the electron, the muonic molecular ion has an internuclear distance that is shorter by the same factor of $\sim200$ compared to that of its electronic counterpart D$_2^+$. There has been great interest on \ddmu\ in connection to studies on Muon Catalyzed Fusion (\mCF), where the short internuclear distance dramatically increases fusion rates with great potential for energy production. From these studies, it was discovered that in some particular cases such as in \ddmu\ and $dt\mu^+$ ions ($t$ for tritium), fusion proceeds with a resonantly-enhanced formation rate with a strong temperature dependence, indicating the existence of weakly bound states. Additional evidence for the resonant formation of the muonic molecules was the observation that the \ddmu\ molecule formation rate strongly depends on the hyperfine structure (HFS) of the $d\mu$ atom, i.e. $F = 3/2$ or $F = 1/2$ hyperfine state. Detailed analysis showed that the binding energy of the $(v=1,K=1)$ weakly-bound state in \ddmu\ can be determined with rather high accuracy of up to sub-meV from fusion-rate studies~\cite{Scrinzi1993}.

Muonic deuterium atoms $d\mu$ are formed when muons are stopped within a D$_2$ sample. When the $(d\mu)_F$ atom scatters off D$_2$ at low energies, with the incident energy $\varepsilon_{in}(d\mu)$ in the order of $0.01$ eV, the \ddmu\ molecular ion is formed via the following resonant mechanism:
\begin{equation}
(d\mu)_F+(\mathrm{D}_2)_{K_iv_i} \to
   \left[(dd\mu)_{11}^{SJ_{t}}dee\right]_{K_fv_f}
\end{equation}
Here $\mathbf{F}=\mathbf{I}_d+\mathbf{s}_\mu$ is the total spin of the $d\mu$ atom, $\mathbf{J}_t = \mathbf{I}+\mathbf{s}_\mu+\mathbf{J}$ is the total angular momentum of the $dd\mu_{11}$ ion; $(K_i,v_i)$ --- initial rotational and vibrational quantum numbers of D$_2^+$ and $(K_f,v_f)$ are those of the final state: $\left[(dd\mu)_{11}^{SJ_{t}}dee\right]$. The latter 6-body muonic molecular deuterium complex is referred as MD denoting a molecular reaction complex with a compound nucleus $dd\mu(1,1)$.

Energy conservation determines the resonance condition:
\begin{equation}\label{r-cond}
\begin{array}{@{}l}\displaystyle
\varepsilon_{in}(d\mu) =
   E_{dd\mu}(11;SJ_t)+E_{{\rm MD}}(K_f,v_f)+\Delta E_{mol-FS}
\\[3mm]\displaystyle\hspace{25mm}
   -E_{d\mu}(F)-E_{{\rm D}_2}(K_i,v_i).
\label{eq:resonance}
\end{array}
\end{equation}
Thus, tuning the $d\mu$ kinetic energy $\varepsilon_{in}(d\mu)$ towards this resonance condition by temperature, greatly enhances the production of \ddmu, and eventually increases the fusion rate.
For the calculation of $E_{{\rm MD}}$ in Eq.~\ref{eq:resonance}, the MD complex is treated as a molecule with two point-like structureless nuclei.
All corrections connected with finite size of the \ddmu\ ion, as well as its spin-dependent interaction within the MD molecular complex (dependent on $J_t$) are incorporated in the $\Delta E_{mol-FS}$ term.
Making use of the resonance condition, the energy of the weakly-bound state of (\ddmu)$_{11}^{SJ_t}$ may be extracted from experimental data, provided that all other energy terms in Eq.~\ref{eq:resonance} are obtained with sufficient accuracy.
Details of the experiment and the fitting models employed to provide the binding energy of the $(dd\mu^+)_{11}$ molecular ion, are discussed in~\cite{Scrinzi1993,Balin2011}. The most recent and comprehensive analysis of Balin \et~\cite{Balin2011} gives the experimental value for the binding energy of $E_\mathrm{exp}(dd\mu^+)_{11} = 1.9651(7)$ eV.

The theoretical binding energies for the $(dd\mu)_{11}^{SJ_t}$ ion were calculated in \cite{Bakalov2001}.
The hyperfineless energy were calculated taking into account leading-order relativistic and radiative corrections, where of particular importance are the vacuum polarization effects, while the hyperfine splitting is treated separately.
In this treatment, the most accurate nonrelativistic calculations from Ref.~\cite{Korobov2004} were employed. In addition to accurate energies with sub-meV precisions, accurate wavefunctions, e.g. those plotted in Fig.~\ref{fig:wfn}, were obtained from these calculations. The finite size of the nuclear charge distribution and the nuclear polarizability have also been included in the theoretical treatment. The final uncertainty of the theoretical binding and HFS energies does not exceed 0.1 meV. It is noted that strong and weak interaction effects at this level do not play a significant role even at this picometer-separations.

The dissociation (or binding) energy is defined as the energy difference between the weakly bound state $(v=1,K=1)$ of the \ddmu\ ion and the state when the two constituent deuterons are non-interacting, i.e. at $r=\infty$ where $V(\infty)=0$. The expectation value of the fifth force for the $(v=1, K=1)$ \ddmu\ state with the wavefunction $\Psi_{(1,1)}$ can be written as
\begin{eqnarray}
	\Braket{\Delta V_{5,\lambda}} = - \alpha_5 N_1 N_2 & \Braket{\Psi_{(1,1)}(r)  | Y(r,\lambda) | \Psi_{(1,1)}(r)} \hbar c. \nonumber \\
\label{Ydiss}
\end{eqnarray}
A comparison of the most recent experimental determination of the $dd\mu^+(1,1)$ binding energy in Ref.~\cite{Balin2011} and the most accurate theoretical results~\cite{Bakalov2001, Korobov2004} demonstrate good agreement within an uncertainty limited by the experiment.
Using the combined experimental-theoretical uncertainty of $\delta E=7 \times 10^{-4}$ eV, bounds were obtained from Eq.~\ref{Eq:constraint} as a function of interaction lengths and plotted in Fig.~\ref{expectation} for $\lambda = 0.004-0.04$ \AA. A constraint of $\alpha_5 < 5 \times 10^{-6} \alpha_{EM}$ is obtained from \ddmu\ for interaction lengths $\lambda > 0.04$ \AA, however, this is superseded by the tighter limits from antiprotonic helium at this interaction range.

\section{Discussion and Conclusion}

Excellent agreement between the experimental and theoretical values for the transition energies of antiprotonic helium and for the binding energy of muonic molecular deuterium ion are shown in the present comparison. The combined experimental-theoretical uncertainty then places an upper limit to effects of unknown fifth-force interactions beyond QED or the Standard model. The general characteristic of these exotic species is the substitution of the electron by a heavier particle, either by an antiproton or muon, leading to the reduction in the distance between the heavy particles.
The resulting interhadronic separation in the sub-\r{A}ngstrom range for these exotic systems enables the extension of the constraints obtained from hydrogenic molecules and ions~\cite{Salumbides2013} to even shorter fifth-force interaction range. 

Previous limits based on antiprotonic helium have been derived based on approximate theory~\cite{Pokotilovski2006, Nesvizhevsky2008}. In contrast, the most advanced \emph{ab initio} theory~\cite{Korobov2008b}, which provides both energies and wavefunctions, is used in this study.
In addition, the fifth force interaction range $\lambda$ is taken as a parameter in the calculation of the expectation values to separate its effect with the interaction strength $\alpha_5$.
The most accurate experimental values of antiprotonic helium transitions by Hori \et~\cite{Hori2011} were used in this comparison. In the discussions up to now, we have treated the antiproton as a hadron with no reference to its anti-matter character. In fact, Hori \et~\cite{Hori2011} established the equivalence of protons and antiprotons within their experimental accuracy, thus our treatment is justified. The present limit obtained from antiprotonic helium therefore covers extra matter-antimatter hadronic interactions.

Constraints obtained from the \ddmu\ molecular ion are obtained here for the first time. The short internuclear distance of this exotic molecule provides relatively stringent limits on the picometer scale.
The present constraint is limited by the experimental determination of the binding energy of the $dd\mu^+(1,1)$ state, obtained from investigations of the resonant nature in the formation-rate of fusion products.
The interpretation of the experimental results relies on theoretical models of the fusion reaction process. It turns out that because of the resonant nature of the fusion process, the rate is very sensitive to $E((dd\mu^+)_{11})$ binding energy such that this parameter can be obtained with high accuracy from the fit.    
An advantage of the method employing a comparison of \emph{ab initio} theory and spectroscopic data as discussed in this letter is its simplicity. This leads to direct interpretation with minimal assumptions, providing constraints that should be robust. Future improvements in the constraints are expected as spectroscopic experiments as well as theory advances. The interparticle distances in atoms, molecules and their exotic counterparts occupy a distance scale that complement the range accessed in other investigations.

\begin{figure}
\includegraphics[width=\columnwidth]{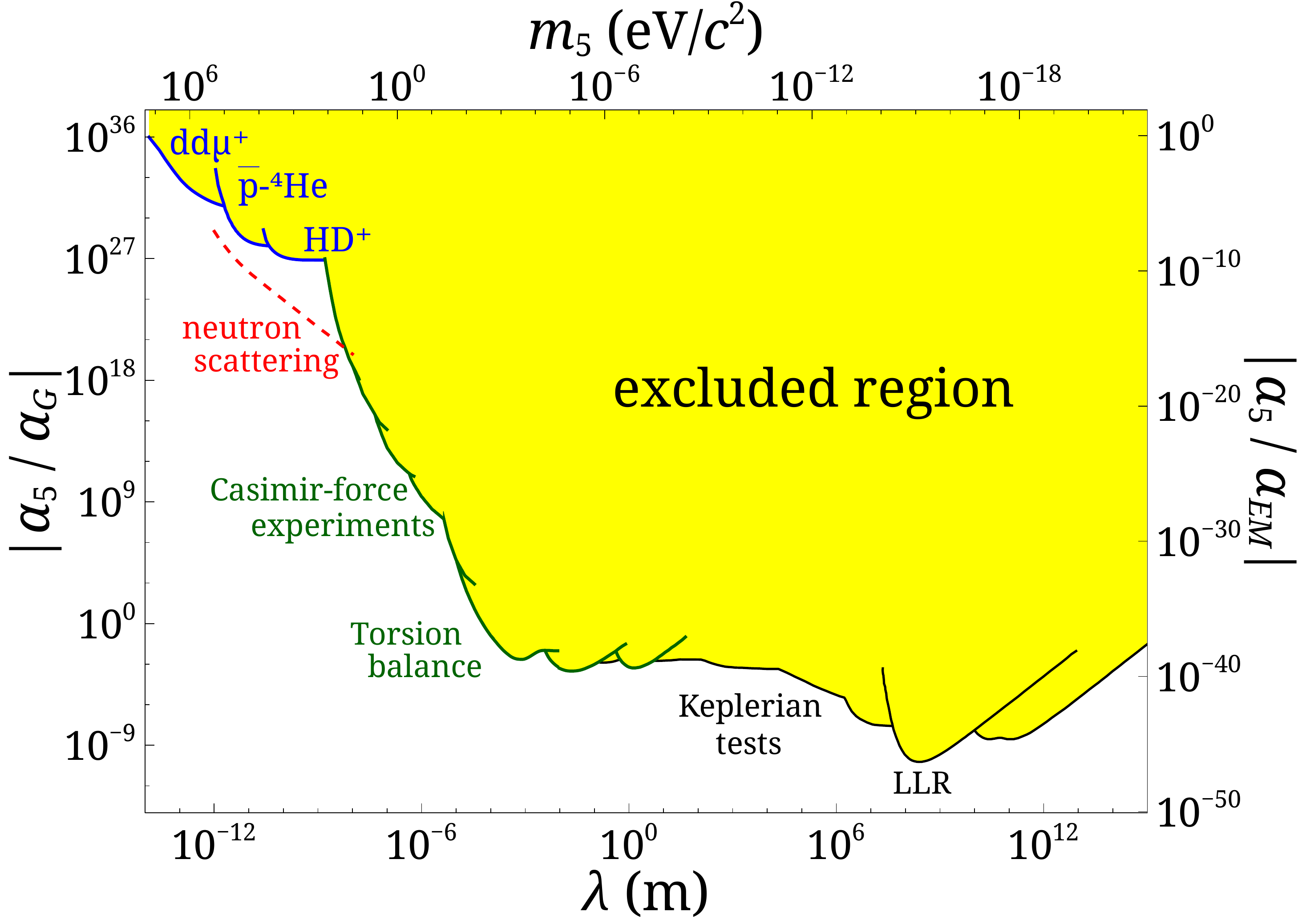}
\caption{
(Color online) Global constraints for $\alpha_5$ for a range of interactions lengths $\lambda$. The figure is based on the present results, those of Ref.~\cite{Salumbides2013}, and on data presented in Ref.~\cite{Adelberger2009} including a wide variety of experiments of different nature. (Keplerian tests include planetary and satellite interactions; LLR: Lunar Laser Ranging)
\label{Fig-Grandpicture}}
\end{figure}

The present bounds in the sub-\r{A}ngstrom scale extend the constraints obtained from inverse-square-law tests of gravity, e.g. those discussed in a recent review~\cite{Antoniadis2011}.
A global overview of the constraints is shown in Fig.~\ref{Fig-Grandpicture}, where $\lambda$ interaction range encompasses some 28 orders of magnitude.
The figure is based on the results presented in Ref.~\cite{Adelberger2009} that included a wide variety of experiments of different nature, and extended towards the sub-atomic regime with the present results. 
It is clear from this picture that the bounds towards shorter interaction ranges quickly become less tight.
On the other hand, the constraints at large $\lambda$ are orders of magnitude more tight, since these are obtained mostly using macroscopic experiments where massive numbers of interacting particles ($N_{1,2}$) are involved. 
The constraints obtained from experiments based on Casimir-force investigations are taken from Adelberger et al.~\cite{Adelberger2009} and references therein. Neutron scattering experiments overlap in the $\lambda$-range accessed in this study, where tighter bounds are claimed~\cite{Nesvizhevsky2008}. However, the analysis undertaken in order to provide constraints from neutron scattering experiments is not as direct as in the method we present.

\section*{Acknowledgments}
This research was supported by the Netherlands Foundation for Fundamental Research of Matter (FOM) through the program "Broken Mirrors \& Drifting Constants".

\end{document}